# Seven temperate terrestrial planets around the nearby ultracool dwarf star TRAPPIST-1

Michaël Gillon[1], Amaury H. M. J. Triaud[2], Brice-Olivier Demory[3,4], Emmanuël Jehin[1], Eric Agol[5,6], Katherine M. Deck[7], Susan M. Lederer[8], Julien de Wit[9], Artem Burdanov[1], James G. Ingalls[10], Emeline Bolmont[11,12], Jeremy Leconte[13], Sean N. Raymond[13], Franck Selsis[13], Martin Turbet[14], Khalid Barkaoui[15], Adam Burgasser[16], Matthew R. Burleigh[17], Sean J. Carey[10], Aleksander Chaushev[17], Chris M. Copperwheat[18], Laetitia Delrez[1,4], Catarina S. Fernandes[1], Daniel L. Holdsworth[19], Enrico J. Kotze[20], Valérie Van Grootel[1], Yaseen Almleaky[21,22], Zouhair Benkhaldoun[15], Pierre Magain[1], Didier Queloz[4,23]

[1]Space sciences, Technologies and Astrophysics Research (STAR) Institute, Université de Liège, Allée du 6 Août 17, Bat. B5C, 4000 Liège, Belgium

[2]Institute of Astronomy, Madingley Road, Cambridge CB3 0HA, UK

[3]University of Bern, Center for Space and Habitability, Sidlerstrasse 5, CH-3012, Bern, Switzerland

[4]Cavendish Laboratory, J J Thomson Avenue, Cambridge, CB3 0HE, UK

[5]Astronomy Department, University of Washington, Seattle, WA, 98195, USA

[6]NASA Astrobiology Institute's Virtual Planetary Laboratory, Seattle, WA, 98195, USA

[7]Department of Geological and Planetary Sciences, California Institute of Technology, Pasadena, CA, USA

[8]NASA Johnson Space Center, 2101 NASA Parkway, Houston, Texas, 77058, USA

[9]Department of Earth, Atmospheric and Planetary Sciences, Massachusetts Institute of Technology, 77 Massachusetts Avenue, Cambridge, MA 02139, USA

[10]Spitzer Science Center, California Institute of Technology, 1200 E California Boulevard, Mail Code 314-6, Pasadena, CA 91125, USA

[11]NaXys, Department of Mathematics, University of Namur, 8 Rempart de la Vierge, 5000 Namur, Belgium

[12]Laboratoire AIM Paris-Saclay, CEA/DRF - CNRS - Univ. Paris Diderot - IRFU/SAp, Centre de Saclay, F- 91191 Gif-sur-Yvette Cedex, France

[13]Laboratoire d'astrophysique de Bordeaux, Univ. Bordeaux, CNRS, B18N, Allée Geoffroy Saint-Hilaire, F-33615 Pessac, France

[14]Laboratoire de Météorologie Dynamique, Sorbonne Universités, UPMC Univ Paris 06, CNRS, 4 place Jussieu, 75005 Paris, France

[15]Laboratoire LPHEA, Oukaimeden Observatory, Cadi Ayyad University/FSSM, BP 2390, Marrakesh, Morocco

[16]Center for Astrophysics and Space Science, University of California San Diego, La Jolla, CA, 92093, USA

[17]Leicester Institute for Space and Earth Observation, Dept. of Physics and Astronomy, University of Leicester, Leicester LE1 7RH, UK

[18]Astrophysics Research Institute, Liverpool John Moores University, Liverpool L3 5RF, UK

[19]Jeremiah Horrocks Institute, University of Central Lancashire, Preston PR1 2HE, UK

[20]South African Astronomical Observatory, PO Box 9, Observatory 7935, Cape Town, South Africa

[21]Space and Astronomy Department, Faculty of Science, King Abdulaziz University, 21589 Jeddah, Saudi Arabia.

[22]King Abdulah Centre for Crescent Observations and Astronomy (KACCOA), Makkah Clock, Saudia Arabia.

[23]Observatoire de Genève, Université de Genève, 51 chemin des Maillettes, CH-1290 Sauverny, Switzerland.

**One focus of modern astronomy is to detect temperate terrestrial exoplanets well-suited for atmospheric characterisation. A milestone was recently achieved with the detection of three Earth-sized planets transiting (i.e. passing in front of) a star just 8% the mass of the Sun 12 parsecs away**[1]**. Indeed, the transiting configuration of these planets combined**

with the Jupiter-like size of their host star - named TRAPPIST-1 - makes possible in-depth studies of their atmospheric properties with current and future astronomical facilities[1,2,3]. Here we report the results of an intensive photometric monitoring campaign of that star from the ground and with the Spitzer Space Telescope. Our observations reveal that at least seven planets with sizes and masses similar to the Earth revolve around TRAPPIST-1. The six inner planets form a near-resonant chain such that their orbital periods (1.51, 2.42, 4.04, 6.06, 9.21, 12.35 days) are near ratios of small integers. This architecture suggests that the planets formed farther from the star and migrated inward[4,5]. The seven planets have equilibrium temperatures low enough to make possible liquid water on their surfaces[6,7,8].

Among the three initially reported TRAPPIST-1 planets, one of them - called 'TRAPPIST-1d' in the discovery publication[1] - was identified based on only two transit signals observed at moderate signal-to-noise. We also observed its second transit signal - blended with a transit of planet c - with the HAWK-I infrared imager on the Very Large Telescope (Chile). Our analysis of the VLT/HAWK-I data - subsequent to the submission of the discovery paper - resulted in a light curve of high enough precision to firmly reveal the triple nature of the observed eclipse (Extended Data Fig. 1). This intriguing result motivated us to intensify our photometric follow-up of the star which resumed in February and March 2016 with observations of six possible transit windows of 'TRAPPIST-1d' with the Spitzer Space Telescope. It continued in May 2016 with the intense ground-based observations of the star with TRAPPIST-South in Chile, its newly-commissioned Northern twin TRAPPIST-North in Morocco, the 3.8m UKIRT telescope at Hawaii, the 4m William Herschel and the 2m Liverpool telescopes at La Palma, and the SAAO 1.0m telescope in South Africa. It culminated on 19 September 2016 with the start of a 20d-long nearly continuous monitoring campaign of the star by the Spitzer Space Telescope at 4.5 μm.

The light curves obtained prior to 19 September 2016 allowed us to discard the eleven possible periods of 'TRAPPIST-1d'[1], indicating that the two observed transits originated from different objects. Furthermore, these light curves showed several transit-like signals of unknown origins that we could not relate to a single period (Extended Data Fig. 2 and 3). The situation was resolved through the 20d-long photometric monitoring campaign of the star by Spitzer. Its resulting light curve shows 34 clear transits (Fig. 1) that - when combined with the ground-based dataset - allowed us to unambiguously identify four periodic transit signals of periods 4.04d, 6.06d, 8.1d and 12.3d that correspond to four new transiting planets named respectively TRAPPIST-1d, e, f, and g (Fig. 1, Extended Data Fig. 2 and 3). The uniqueness of the solution is ensured by the sufficient numbers of unique transits observed per planet (Table 1), by their consistent shapes for each planet (see below), and by the near-continuous nature of the Spitzer light curve and its duration longer than the periods of the four planets. The Spitzer photometry also shows an orphan transit-shaped signal with a depth of ~0.35% and a duration of ~75min occurring at JD~2,457,662.55 (Fig. 1) that we attribute to a seventh outermost planet of unknown orbital period, TRAPPIST-1h. We combed our ground-based photometry in search of a second transit of this planet h, but no convincing match was identified.

We analysed our extensive photometric dataset in three phases. First, we performed individual analyses of all transit light curves with an adaptive Markov-Chain Monte Carlo (MCMC) code[1,9] to measure their depths, durations, and timings (see Methods). We derived a mean transit ephemeris for each planet from their measured transit timings. We successfully checked the consistency of the durations and depths of the transits for planets b to g. For each planet, and especially for f and g, the residuals of the fit show transit timing variations (TTVs) with amplitudes ranging from a few tens of seconds to more than 30 minutes that indicate significant mutual interactions between the planets[10,11,12] (Extended Fig. 2 and 3).

In a second phase, we performed a global MCMC analysis of the transits observed by Spitzer to constrain the orbital and physical parameters of the seven planets. We decided to use only the Spitzer data due to their better precision compared with most of our ground-based data, and of the minimal amplitude of the limb-darkening at 4.5μm which strengthens constraints possible on the transit shapes - and thus on the stellar density and, by extension, on the physical and orbital parameters of the planets[13]. We assumed circular orbits for all of the planets, based on the results of N-body dynamical simulations that predicted orbital eccentricities < 0.1 for the six inner planets (Table 1); the orbital eccentricity of the outer planet h cannot be constrained from a single transit. This global analysis assumed the *a priori* knowledge for the star that is described in ref. 1 (see Methods). To account for significant planet-planet interaction, TTVs were included as free parameters for the six inner planets. We used each planet's transit ephemeris (derived in the first phase) as a prior on the orbital solution.

In a third phase, we used the results obtained above to investigate the TTV signals themselves. By performing a series of analytical and numerical N-body integrations (see Methods), we could determine initial mass estimates for the six inner planets, along with their orbital eccentricities. We emphasise the preliminary nature of this dynamical solution which may not correspond to a global minimum of the parameter space, and that additional transit observations of the system will be required to lift the existing degeneracies (see Methods).

Table 1 shows the main planetary parameters derived from our data analysis. We find that five planets (b, c, e, f, g) have sizes similar to the Earth, while the other two (d and h) are intermediate between Mars (~0.5 $R_{Earth}$) and Earth. The mass estimates for the six inner planets broadly suggest rocky compositions[14] (Fig. 2.a). Their precisions are not high enough to constrain the fraction of volatiles in the planets' compositions, except for planet f whose low density favors a volatile-rich composition. The volatile content could be in the form of an ice layer and/or of an atmosphere, something that can be verified with follow-up observations during transit with space telescopes like Hubble[2] and James Webb[3]. We note that the ratio of masses between the six inner planets and TRAPPIST-1 and that of the Galilean satellites and Jupiter are both ~0.02%, maybe implying a similar formation history[15,16].

The derived planets' orbital inclinations are all very close to 90°, indicating a dramatically coplanar system seen nearly edge-on. Furthermore, the six inner planets form the longest currently-known near-resonant chain of exoplanets, with the orbital periods ratios $P_c/P_b$, $P_d/P_c$, $P_e/P_d$, $P_f/P_e$, and $P_g/P_f$ close to the ratios of small integers 8:5, 5:3, 3:2, 3:2, and 4:3, respectively. This proximity to mean motion resonances of several planet pairs explains the significant amplitudes of the measured TTVs. Similar near-resonant chains involving up to four planets have been discovered in compact systems containing super-Earths and Neptunes orbiting Sun-like stars[5,17]. Orbital resonances are naturally generated when multiple planets interact within their nascent gaseous discs[18]. The favoured theoretical scenario for the origin of the TRAPPIST-1 system is an accretion of the planets farther from the star followed by a phase of disc-driven inward migration[4,19], a process first studied in the context of the Galilean moons around Jupiter[20]. The planets' compositions should reflect their formation zone so this scenario predicts that the planets should be volatile-rich and have lower densities than Earth[21,22], in good agreement with our preliminary result for TRAPPIST-1f (Fig. 2.a).

The stellar irradiation of the planets cover a range of ~4.3 to ~0.13 $S_{Earth}$ (=solar irradiation at 1 au) which is very similar to the one of the inner solar system (Mercury=6.7 $S_{Earth}$, Ceres=0.13 $S_{Earth}$,). Notably, planets c, d, and f have stellar irradiations very close to those of Venus, Earth, and Mars, respectively (Fig. 2). However, even at these low insolations, all seven planets are expected to be either tidally synchronized[23], or trapped in a higher-order spin-orbit resonance, the latter being rather unlikely considering the constraints on the orbital eccentricities[24] (Table 1). Using a 1D cloud-free climate model accounting for the low-temperature spectrum of the host star[25], we infer that the three planets e, f, and g could harbour water oceans on their surfaces, assuming Earth-like atmospheres. The same inference is obtained when running a 3D climate model[26] assuming that the planets are tidally synchronous. For the three inner planets (b,c,d), our 3D climate modeling results in runaway greenhouses. The cloud feedback that usually decreases the surface temperatures for synchronous planets is rather inefficient for such short period objects[27]. Nevertheless, if some water survived the hot early phase of the system[28], the irradiation received by planets (b,c,d) are still low enough to make possible for limited regions on their surfaces to harbour liquid water[1,7]. While the orbital period, and therefore distance of planet h is not yet well defined, its irradiation is probably too low to sustain surface temperatures above the melting point of water. However, it could still harbour surface liquid water providing a large enough internal energy - e.g. from tidal heating - or the survival of a significant fraction of its primordial $H_2$-rich atmosphere that could strongly slow down the loss of its internal heat[8].

We found the long-term dynamical evolution of the system to be very dependent on the exact orbital parameters and masses of the seven planets, which are currently too uncertain to make possible any reliable prediction (see Methods). All our dynamical simulations predict small but non-zero orbital eccentricities for the six inner planets (see 2-σ upper limits in Table 1). The resulting tidal heating could be strong enough to significantly impact their energy budgets and geological activities[29].

The TRAPPIST-1 system is a compact analog of the inner solar system (Fig. 2.b). It represents a unique opportunity to thoroughly characterise[1,2,3] temperate Earth-like planets orbiting a much cooler and smaller star than the Sun, and notably to study the impact of tidal locking[22], tidal heating[29], stellar activity[22] and an extended pre-main-sequence phase[30] on their atmospheric properties.

**Author Contributions.** MG leads the ultracool dwarf transit survey of TRAPPIST and the photometric follow-up of TRAPPIST-1, planned and analysed most of the observations, led their scientific exploitation, and wrote most of the manuscript. AHMJT led the observational campaign with La Palma telescopes (LT & WHT), CMC managed the scheduling of the LT observations, and ArB performed the photometric analysis of the resulting LT & WHT images. BOD led the TTV/dynamical simulations. EA and KMD performed independent analyses of the transit timings. JI and SC helped optimizing the Spitzer observations. BoD, JI, and JdW performed independent analyses of the Spitzer data. MG, EJ, LD, ArB, PM, KB, YA, and ZB performed the TRAPPIST observations and their analysis. SL obtained the DD time on UKIRT and managed with EJ the preparation of the UKIRT observations. MT, JL, FS, EB, and SNR performed atmospheric modeling for the planets and worked on the theoretical interpretation of their properties. VVG managed the SAAO observations performed by CSF, MRB, DLH, AS and EJK. All co-authors assisted writing the manuscript. AHMJT prepared most of the figures in the paper.

**Author Information.** Reprints and permissions information is available at www.nature.com/reprints. The authors declare no competing financial interests. Readers are welcome to comment on the online version of the paper. Correspondence and requests for materials should be addressed to M.G. (michael.gillon@ulg.ac.be).

**Online Content.** Methods, along with any additional Extended Data display items and Source Data, are available in the online version of the paper; references unique to these sections appear only in the online paper. Acknowledgments are presented online as Supplementary Information.

**Table 1 | Updated properties of the TRAPPIST-1 planetary system**

| Parameter | Value | | | | | | |
|---|---|---|---|---|---|---|---|
| **Star** | **TRAPPIST-1 = 2MASS J23062928-0502285** | | | | | | |
| Magnitudes[1] | V=18.8, R=16.6, I=14.0, J=11.4, K=10.3 | | | | | | |
| Distance [pc][1] | 12.1±0.4 | | | | | | |
| Mass $M_\star$ [$M_\odot$][a] | 0.0802±0.0073 | | | | | | |
| Radius $R_\star$ [$R_\odot$][a] | 0.117±0.0036 | | | | | | |
| Density $\rho_\star$ [$\rho_\odot$] | $50.7^{+1.2}_{-2.2}\,\rho_\odot$ | | | | | | |
| Luminosity $L_\star$ [$L_\odot$][a] | 0.000524±0.000034 | | | | | | |
| Effective temperature $T_{eff}$ [K][a] | 2559±50 | | | | | | |
| Metallicity [Fe/H][a] [dex] | +0.04±0.08 | | | | | | |
| **Planets** | **b** | **c** | **d** | **e** | **f** | **g** | **h** |
| Number of unique transits observed | 37 | 29 | 9 | 7 | 4 | 5 | 1 |
| Period $P$ [d] | 1.51087081 ±0.60×10⁻⁶ | 2.4218233 ±0.17×10⁻⁵ | 4.049610 ±0.63×10⁻⁴ | 6.099615 ±0.11×10⁻⁴ | 9.206690 ±0.15×10⁻⁴ | 12.35294 ±0.12×10⁻³ | $20^{+15}_{-6}$ |
| Mid-transit time $T_0$ - 2,450,000 [BJD$_{TDB}$] | 7322.51736 ±0.00010 | 7282.80728 ±0.00019 | 7670.14165 ±0.00035 | 7660.37859 ±0.00038 | 7671.39767 ±0.00023 | 7665.34937 ±0.00021 | 7662.55463 ±0.00056 |
| Transit depth $(R_p/R_\star)^2$ [%] | 0.7266 ±0.0088 | 0.687 ±0.010 | 0.367 ±0.017 | 0.519 ±0.026 | 0.673 ±0.023 | 0.782 ±0.027 | 0.352 ±0.0326 |
| Transit impact parameter $b$ [$R_\star$] | $0.126^{+0.092}_{-0.078}$ | $0.161^{+0.076}_{-0.084}$ | 0.17±0.11 | $0.12^{+0.11}_{-0.09}$ | 0.382 ±0.035 | 0.421 ±0.031 | $0.45^{+0.22}_{-0.29}$ |
| Transit duration $W$ [min] | 36.40±0.17 | 42.37±0.22 | 49.13±0.65 | 57.21±0.71 | 62.60±0.60 | 68.40±0.66 | $76.7^{+2.7}_{-2.0}$ |
| Inclination $i$ [°] | $89.65^{+0.22}_{-0.27}$ | 89.67±0.17 | 89.75±0.16 | $89.86^{+0.10}_{-0.12}$ | 89.680 ±0.034 | 89.710 ±0.025 | $89.80^{+0.10}_{-0.05}$ |
| Eccentricity $e$ (2-σ upper limit from TTVs) | <0.081 | <0.083 | <0.070 | <0.085 | <0.063 | <0.061 | - |
| Semi-major axis $a$ [10⁻³ au] | 11.11±0.34 | 15.21±0.47 | $21.44^{+0.66}_{-0.63}$ | $28.17^{+0.83}_{-0.87}$ | 37.1±1.1 | 45.1±1.4 | $63^{+27}_{-13}$ |
| Scale parameter $a/R_\star$ | $20.50^{+0.16}_{-0.31}$ | $28.08^{+0.22}_{-0.42}$ | $39.55^{+0.30}_{-0.59}$ | $51.97^{+0.40}_{-0.77}$ | $68.4^{+0.5}_{-1.0}$ | $83.2^{+0.6}_{-1.2}$ | $117^{+50}_{-26}$ |
| Irradiation $S_p$ [$S_{Earth}$] | 4.25±0.33 | 2.27±0.18 | 1.143 ±0.088 | 0.662 ±0.051 | 0.382 ±0.030 | 0.258 ±0.020 | $0.131^{+0.081}_{-0.067}$ |
| Equilibrium temperature [K][b] | 400.1 ±7.7 | 341.9 ±6.6 | 288.0 ±5.6 | 251.3 ±4.9 | 219.0 ±4.2 | 198.6 ±3.8 | $168^{+21}_{-28}$ |
| Radius $R_p$ [$R_{Earth}$] | 1.086 ±0.035 | 1.056 ±0.035 | 0.772 ±0.030 | 0.918 ±0.039 | 1.045 ±0.038 | 1.127 ±0.041 | 0.755 ±0.034 |
| Mass $M_p$ [$M_{Earth}$] (from TTVs) | 0.85 ±0.72 | 1.38 ±0.61 | 0.41 ±0.27 | 0.62 ±0.58 | 0.68 ±0.18 | 1.34 ±0.88 | - |
| Density $\rho_p$ [$\rho_{Earth}$] | 0.66 ±0.56 | 1.17 ±0.53 | 0.89 ±0.60 | 0.80 ±0.76 | 0.60 ±0.17 | 0.94 ±0.63 | - |

The values and 1-sigma errors for the parameters of TRAPPIST-1 and its seven planets, as deduced for most parameters from a global analysis of the Spitzer photometry, including *a priori* knowledge on the stellar properties. Masses of the planets and upper limits on their eccentricities were deduced from the analysis of the TTVs (see text and Methods). We outline that the planet TRAPPIST-1d does not correspond to the discarded 'TRAPPIST-1d' candidate presented in ref. 1 (see text).

[a]Informative prior probability distribution functions were assumed for these stellar parameters (see Methods).

[b]Assuming a null Bond albedo.

**Figure 1 | The TRAPPIST-1 system as seen by Spitzer.** *a* and *b*. Spitzer photometric measurements (dark points) resulting from the nearly-continuous observation of the star from 19 September to 10 October 2016. The ground-based measurements (binned per 5 min for clarity) gathered during the Spitzer gaps are shown as light grey points. The position of the transits of the planets are shown as coloured diamonds. *c.* Period-folded photometric measurements obtained by Spitzer near transits of planets TRAPPIST-1b-h corrected for the measured TTVs. Coloured dots show the unbinned measurements, whereas the open circle depict binned measurements for visual clarity. The 1-sigma error bars of the binned measurements are shown as vertical lines. The best-fit transit models are shown as coloured lines. 16-11-5-2-3-2-1 transits were observed by Spitzer and combined to produce the shown light curves for planets b-c-d-e-f-g-h, respectively. *d.* Representation of the orbits of the 7 planets. The same colour code as in the two other panels is used to identify the planets. The grey annulus and the two dashed lines represent the zone around the star where abundant long-lived liquid water (i.e. oceans) could exist on the surfaces of Earth-like planets as estimated under two different assumptions in ref. 6. The relative positions of the planets corresponds to their orbital phase during the first transit we detected on this star, by TRAPPIST-1c (the observer is located on the right hand-side of the plot).

**Figure 2 | Mass-radius and incident flux-radius diagrams for terrestrial planets.** In both panels, the coloured circular symbols are the TRAPPIST-1 planets, and the horizontal and vertical lines are 1-sigma error bars. *a.* Mass-radius relation for planets between 0.5 and 1.5 Earth radii, and between 0.1 and 2 Earth masses. The solid lines are theoretical mass-radius curves[14] for planets with different compositions. The fiducial model is 100% $MgSiO_3$ (rock), whose fractional part is decreasing either with increasing fraction of water (the radius increases), or with increasing fractions of Iron (the radius decreases). *b.* Radius *vs* incident flux. Venus and Earth are shown as grey circular symbols, and Mercury, Mars, and Ceres as dotted vertical lines. The planet h has large errors on its irradiation because of its unknown orbital period.

## METHODS

**Observations and photometry**
In addition to the ground-based observations described in ref. 1, this work was based on 1333 hrs of new observations gathered from the ground with the 60cm telescopes TRAPPIST-South (469 hrs) and TRAPPIST-North (202 hrs), the 8m Very Large Telescope (3 hrs), the 4.2m William Herschel telescope (26 hrs), the 4m UKIRT telescope (25 hrs), the 2m Liverpool telescope (50 hrs), and the 1m SAAO telescope (11 hrs), and from space with Spitzer (518 hrs).

The new observations of the star gathered by the TRAPPIST-South[1,31,32] 60cm telescope (La Silla Observatory, Chile) occurred on the nights of 29 to 31 December 2015 and from 30 April to 11 October 2016. The observational strategy used was the same as that described in ref. 1 for previous TRAPPIST-South observations of the star.

TRAPPIST-North[33] is a new 60cm robotic telescope installed in spring 2016 at Oukaïmeden Observatory in Morocco. It is an instrumental project led by the University of Liège, in collaboration with the Cadi Ayyad University of Marrakesh, that is, like its southern twin TRAPPIST-South, totally dedicated to the observations of exoplanet transits and small bodies of the solar system. TRAPPIST-North observations of TRAPPIST-1 were performed from 1 June to 12 October 2016. Each run of observations consisted of 50s exposures obtained with a thermoelectrically-cooled 2k×2k deep-depletion CCD camera (field of view of 19.8' × 19.8', image scale of 0.61"/pixel). The observations employed the same 'I+z' filter as for most of the TRAPPIST-South observations[1].

The new VLT/HAWK-I[34] (Paranal Observatory, Chile) observations that revealed a triple transit of planets c-e-f (see main text and Extended Data Fig. 1) were performed during the night of 10 to 11 December 2015 with the same observational strategy than described in ref. 1 (NB2090 filter), except that each exposure was composed of 18 integrations of 2s.

The 4m telescope UKIRT (Mauna Kea, Hawaii) and its WFCAM infrared camera[35] observed the star on 24 June, 16-18-29-30 July, and 1 August 2016. Here too, the exact same observational strategy as its previous observations of the star[1] was used for these new observations (J filter, exposures of 5 integrations of 1s) .

The 4.2m William Herschel Telescope (La Palma, Canary Islands) observed the star for three nights in a row from 23 to 25 August 2016 with its optical 2k × 4k ACAM camera[36] that has an illuminated circular field of view of 8' diameter and an image scale of 0.25"/pixel. The observations were performed in the Bessel I filter with exposure times between 15 and 23s.

10 runs of observation of TRAPPIST-1 were performed by the robotic 2m Liverpool Telescope between June and October 2016. These observations were obtained through a Sloan-z filter with the 4k × 4k IO:O CCD camera[37] (field of view of 10' × 10'). A 2 × 2 binning scheme resulted in an image scale of 0.30"/pixel. An exposure time of 20s was used for all images.

The 1m telescope at the South African Astronomical Observatory (Sutherland, South Africa) observed the star on the nights of 18-19 June, 21-22 June, and 2-3 July 2016. The observations consisted of 55s exposures taken by the 1k × 1k SHOC CCD camera[38] (field of view of 2.85' × 2.85') using a Sloan z filter and with a 4 × 4 binning, resulting in an image scale of 0.67"/pixel.

For all ground-based data, a standard pre-reduction (bias, dark, flat-field correction) was applied, followed by the measurements of the stellar fluxes from the calibrated images using the DAOPHOT aperture photometry software[39]. In a final stage, a selection of stable comparison stars was manually performed to obtain the most accurate differential photometry possible for TRAPPIST-1.

The Spitzer Space Telescope observed TRAPPIST-1 with its IRAC detector[40] for 5.7 hrs on 21 February 2016, 6.5 hrs on 3-4-7-13-15-18 March 2016, and continuously from 19 September to 10 October 2016. All these observations were done at 4.5 μm in subarray mode (32x32 pixel windowing of the detector) with an exposure time of 1.92s. The observations were done without dithering and in the PCRS peak-up mode[41] that maximizes the accuracy in the position of the target on the detector to minimize the so-called pixel phase effect of IRAC InSb arrays[42]. All the Spitzer data were calibrated with the Spitzer pipeline S19.2.0 and delivered as cubes of 64 subarray images. Our photometric extraction was identical to the one described in ref. 43. DAOPHOT was used to measure the fluxes by aperture photometry and the measurements were combined per cube of 64 images. The photometric errors were taken as the errors on the average flux measurements for each cube.

The observations used in this work are summarized in Extended Data Table 1.

**Analysis of the photometry**
The total photometric dataset - including the data presented in ref. 1 - consists in 81,493 photometric measurements spread over 351 light curves. We converted each universal time (UT) of mid-exposure to the $BJD_{TDB}$ time system[44]. We then performed an individual model selection for each light curve, tested a large range of models composed of a baseline model representing the flux variations correlated to variations of external parameters (e.g. point-spread function size or position on the chip, time, airmass) as low-order (0 to 4) polynomial functions, eventually added to a transit model[45] and/or to a flare model (instantaneous flux increase followed by an exponential decrease) if a structure consistent in shape with these astrophysical signals was visible in the light curve (two of them were captured by Spitzer during its 20d-monitoring campaign, see Fig. 1). The final model of each light curve was selected by minimization of the Bayesian Information Criterion (BIC)[46]. For all the Spitzer light curves, it was necessary to include a linear or quadratic function of the $x$- and $y$-positions of the point-spread function (PSF) centre (as measured in the images by the fit of a 2D-gaussian profile) in the baseline model to account for the pixel phase effect[42,43], complemented for some light curves by a linear or quadratic function of the measured widths of the PSF in the $x$- and/or $y$-directions[43].

For each light curve presenting a transit-like structure whose existence was favoured by the BIC, we explored the posterior probability distribution function (PDF) of its parameters

(width, depth, impact parameter, mid-transit timing) with an adaptive Markov-chain Monte Carlo (MCMC) code[1,9]. For the transits originating from the firmly confirmed planets b and c, we fixed the orbital period to the values presented in ref. 1. For the other transit-like structures, the orbital period was also a free parameter. As in ref. 1, circular orbits were assumed for the planets, and the normal distributions $N(0.04, 0.08^2)$ dex, $N(2,555, 85^2)$ K, $N(0.082, 0.011^2)$ $M_\odot$, and $N(0.114, 0.006^2)$ $R_\odot$ were assumed as prior PDF for the stellar metallicity, effective temperature, mass, and radius, respectively, on the basis of *a priori* knowledge of the stellar properties[47,1]. A quadratic limb-darkening law was assumed for the star[48] with coefficients interpolated for TRAPPIST-1 from the tables of ref. 49. The details of the MCMC analysis of each light curve were the same as described in ref. 1.

The resulting values for the timings of the transits were then used to identify planetary candidates by searching for periodicities and consistency between the derived transit shape parameters. Owing to the high-precision and nearly-continuous nature of the photometry acquired by Spitzer on September and October 2016, this process allowed us to firmly identify the four new planets d-e-f-g with periods of 4.1d, 6.1d, 9.2d and 12.3d (Extended Data Fig. 2 & 3). We then measured updated values for their transit timings through new MCMC analyses of their transit light curves for which the orbital periods were fixed to the determined values. For the six planets b-c-d-e-f-g, we then performed a linear regression analysis of the measured transit timings as a function of their epochs to derive a transit ephemeris $T_i = T_0 (\pm\sigma_{T_0}) + E_i \times P (\pm\sigma_P)$, with $T_0$ the timing of a reference transit for which the epoch is arbitrarily set to 0, $P$ the orbital period, and $\sigma_{T_0}$ and $\sigma_P$ their errors as deduced from the covariance matrix (Table 1). For all planets, the residuals of the fit showed some significant deviation indicating TTVs, which is unsurprising given the compactness of the system and the near-resonant chain formed by the six inner planets (see below).

For a transit-like signal observed by Spitzer at JD~2,457,662.55 (Fig. 1), the significance of the detection (>10σ) was large enough to allow us to conclude that a seventh, outermost planet exists as well. This conclusion is not only based on the high significance of the signal and the consistency of its shape with one expected for a planetary transit, but also on the photometric stability of the star at 4.5 μm (outside of the frequent transits and the rare - about 1 per week - flares) as revealed by Spitzer (Fig. 1).

In a final stage, we performed the global MCMC analysis of the 35 transits observed by Spitzer which is described in the main text. It consisted in 2 chains of 100,000 steps whose convergence was successfully checked using the statistical test of Gelman & Rubin[50]. The parameters derived from this analysis for the star and its planets are shown in Table 1.

**TTV analysis**
We used the TTV method[10,11] to estimate the masses of the TRAPPIST-1 planets. The continuous exchange of angular momentum between gravitationally interacting planets causes them to accelerate and decelerate along their orbits, making their transit times occur early or late compared to a Keplerian orbit[14].

All six inner TRAPPIST-1 planets exhibit transit timing variations due to perturbations from their closest neighbours (Extended Data Fig. 4). The TTV signal for each planet is dominated primarily by interactions with adjacent planets, and these signals have the potential to be particularly large because each planet is near a mean motion resonance with its neighbours. As calculated from the current data, the TTV amplitudes range in magnitude from 2 to more than 30 minutes However, the distances of these pairs to exact resonances controls the amplitude and the period of the TTV signals and is not precisely pinned down by the current dataset. Additionally, the relatively short timeframe during which transits have been monitored prevents an efficient sampling of the TTV oscillation frequencies for the different pairs of planets defined by $f(TTV) = n_i/P_i - n_j/P_j$, where $P$ is the orbital period, $n$ the mean motion, and $i$ and $j$ the planet indices[10].

We modeled TTV using both numerical (TTVFast[51], Mercury[52]) and analytical (TTVFaster[53]) integrations of a system of six gravitationally interacting, coplanar planets. TTVFaster is based on analytic approximations of TTVs derived using perturbation theory and includes all terms at first order in eccentricity. Furthermore, it only includes perturbations to a planet from adjacent planets. To account for the 8:5 and 5:3 near resonances in the system, we also included the dominant terms for these resonances which appear at second and third order in the eccentricities. We determined these higher order terms using the results of ref. 54. TTVFaster has the advantage that the model is significantly faster to compute compared with N-body integrations. It is applicable for this system given the low eccentricities determined via TTV analysis (determined independently from N-body integrations and self-consistently with TTVFaster).

Two different minimization techniques were used: Levenberg-Marquardt[55] and Nelder-Mead[56]. For the purpose of the analyses, we used the 98 independent transit times for all six planets and 5 free parameters per planet (mass, orbital period, transit epoch and eccentricity vectors $e\cos\omega$ and $e\sin\omega$, with $e$ the eccentricity and $\omega$ the argument of periastron). We elected not to include the seventh planet TRAPPIST-1h in the fit because only a single transit has been observed and there is not yet an indication of detectable interactions with any of the inner planets. Likewise, we did not detect any perturbation that would require the inclusion of an additional, undetected non-transiting planet in the dynamical fit. The 6-planet model provided a good fit to the existing data (Extended Data Fig. 4), and we found no compelling evidence for extending the current model complexity given the existing data.

Our three independent analyses of the same set of transit timings revealed multiple, mildly inconsistent, solutions that fit the data equally well provided non-circular orbits are allowed in the fit. It is likely that this solution degeneracy originates from the high-dimensionality of the parameter space combined with the limited constraints brought by the current dataset. The best-fit solution that we found - computed with Mercury - has a chi-squared of 92 for 68 degrees of freedom, but involves non-negligible eccentricities (0.03 to 0.05) for all planets, likely jeopardising the long-term stability of the system. In this context, we decided to present conservative estimates of the planets' masses and upper limits for the eccentricities without favouring one of the three independent analyses. For each parameter, we considered as the 1-σ lower/upper limits the smallest/largest values of the 1-σ lower/upper limits of the three posterior PDFs, and the average of the two computed limits as the most representative value. The values and error bars computed for the planets' masses and the 2-σ upper limits for their orbital eccentricities are given in Table 1.

Additional precise transit timings for all seven planets will be key in constraining further the planet masses and eccentricities and in isolating a unique, well-defined, dynamical solution.

**Preliminary assesment of the long-term stability of the system**
We investigated the long-term evolution of the TRAPPIST-1 system using two N-body integration packages: Mercury[52] and WHFAST[57]. We started from the orbital solution produced in Table 1, and integrated over 0.5 Myr. This corresponds to roughly 100 million orbits for planet b. We repeated this procedure by sampling a number of solutions within the 1-σ intervals of confidence. Most integrations resulted in the disruption of the system on a 0.5 Myr timescale.

We then decided to employ a statistical method yielding the probability for a system to be stable for a given period of time, based on the planets' mutual separations[58]. Using the masses and semi-major axes in Table 1, we calculated the separations between all adjacent pairs of planets in units of their mutual Hill spheres[58]. We found an average separation of 10.5 ± 1.9 (excluding planet h), where the uncertainty is the rms of the six mutual separations. We computed that TRAPPIST-1 has a 25% chance of suffering an instability over 1 Myr, and 8.1% to survive over 1 Gyr, in line with our N-body integrations.

Those results obtained by two different methods imply that the TRAPPIST-1 system could be unstable over relatively short timescales. However, they do not take into account the proximity of the planets to their host star and the resulting strong tidal effects that can act to stabilise the system. We included tidal effects in an ameliorated version of the Mercury package[59,60], and found that they significantly enhance the system's stability. However, the disruption is only postponed by tides in most simulations, and further investigations are needed in order to better understand the dynamics of the system. In general, the stability of the system appears to be very dependent on the assumptions on the orbital parameters and masses of the planets, and on the inclusion or exclusion of planet h and on its assumed orbital period and mass. It is also possible that other, still undetected, planets help stabilizing the system. The masses and exact eccentricities of the planets remain currently uncertain, and our results make likely that only a very small number of orbital configurations lead to stable configurations. For instance, mean-motion resonances can protect planetary systems over long timescales[61]. The system clearly exists, and it is unlikely that we are observing it just before its catastrophic disruption, so it is most probably stable over a significant timescale. These facts and the results of our dynamical simulations indicate that, provided enough data, the very existence of the system should bring strong constraints on its components' properties: masses, orbital elements, tidal dissipation efficiencies, which are dependent on the planets' compositions, mutual tidal effects of the planets, mutual inclinations, orbit of planet h, existence of other, maybe not transiting planets, etc.

**Code availability**
The conversion of the UT times of the photometric measurements to the BJD$_{TDB}$ system was performed using the online program created by J. Eastman and distributed at http://astroutils.astronomy.ohio-state.edu/time/utc2bjd.html. The MCMC software used to analyse the photometric data is a custom Fortran 90 code that can be obtained from the corresponding author on reasonable request. The N-body integration codes TTVFast, TTVFaster, and Mercury are freely available online at https://github.com/kdeck/TTVFast, https://github.com/ericagol/TTVFaster, and https://github.com/smirik/mercury. To realise Fig.2a, we relied on TEPCAT, an online catalogue of transiting planets maintained by John Southworth (http://www.astro.keele.ac.uk/jkt/tepcat/).

**Data availability**

The Spitzer data that support the findings of this study are available from the Spitzer Heritage Archive database (http://sha.ipac.caltech.edu/applications/Spitzer/SHA). Source data for Fig. 1 and Extended Data Fig. 1, 2, 3, and 4 are provided with the paper. The other datasets generated during and/or analysed during the current study are available from the corresponding author on reasonable request.

**Extended Data Table 1 | Summary of the observations set used in this work.** For each facility/instrument, the following parametrs are given: the effective number of observation (not accounting for calibration and overhead times), the year(s) of observation, the number of resulting light curves, the used filter or grism, and the number of transits observed for the seven planets TRAPPIST-1 b-c-d-e-f-g-h.

**Extended Data Figure 1 | Light curve of a triple transit of planets c-e-f**. The black points show the differential photometric measurements extracted from VLT/HAWK-I images, with the formal 1-sigma errors shown as vertical lines. The best-fit triple transit model is shown as a red line. Possible configurations of the planets relative to the stellar disc are shown below the light curve for three different times (red = planet c, yellow = planet e, green = planet f). The relative positions and sizes of the planets, as well as the impact parameters correspond to the values given in Table 1.

**Extended Data Figure 2 | Transit light curve of TRAPPIST-1d and e.** The black points show the photometric measurements - binned per 0.005d = 7.2min. The error for each bin (shown as vertical line) was computed as the 1-sigma error on the average. These light curves are divided by their best-fit instrumental models and by the best-fit transit models of other planets (for multiple transits). The best-fit transit models are shown as solid lines. The light curves are period-folded on the best-fit transit ephemeris given in Table 1, their relative shifts on the *x*-axis reflecting TTVs due to planet-planet interactions (see text). The epoch of the transit and the facility used to observe it are mentionned above each light curve.

**Extended Data Figure 3 |** Transit light curves of TRAPPIST-1f and g. Same as Extended Data Fig. 2 for the planets f and g.

**Extended Data Figure 4 | Transit Timing Variations (TTVs) measured for TRAPPIST-1b-c-d-e-f-g.** For each planet, the best-fit TTV model computed with the N-body numerical integration code Mercury[52] is shown as a red line. The 1-sigma errors of the transit timing measurements are show as vertical lines.

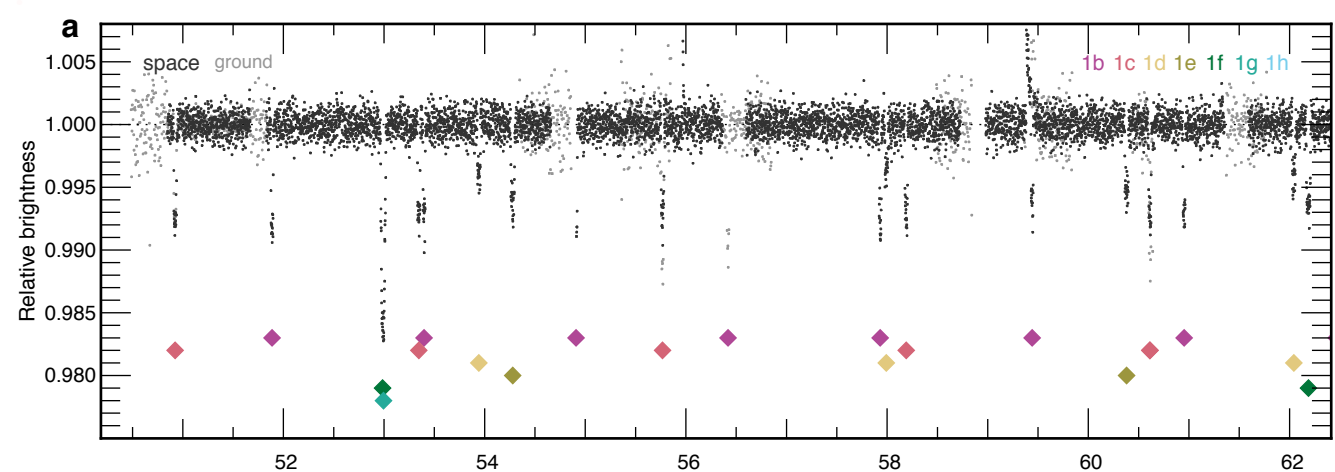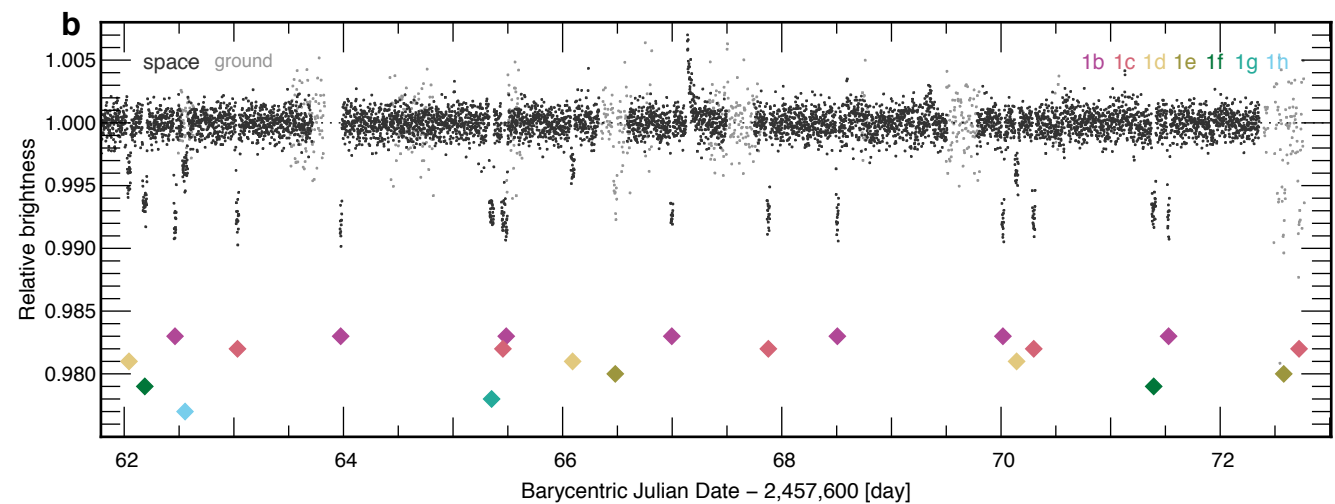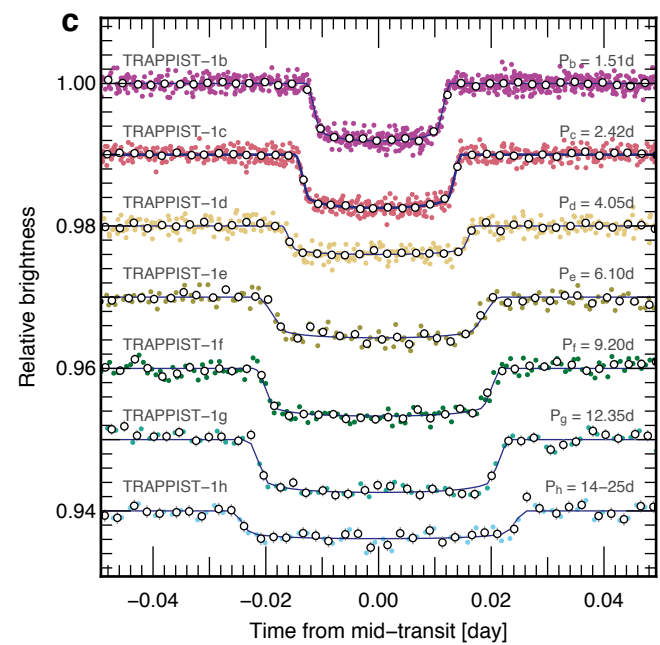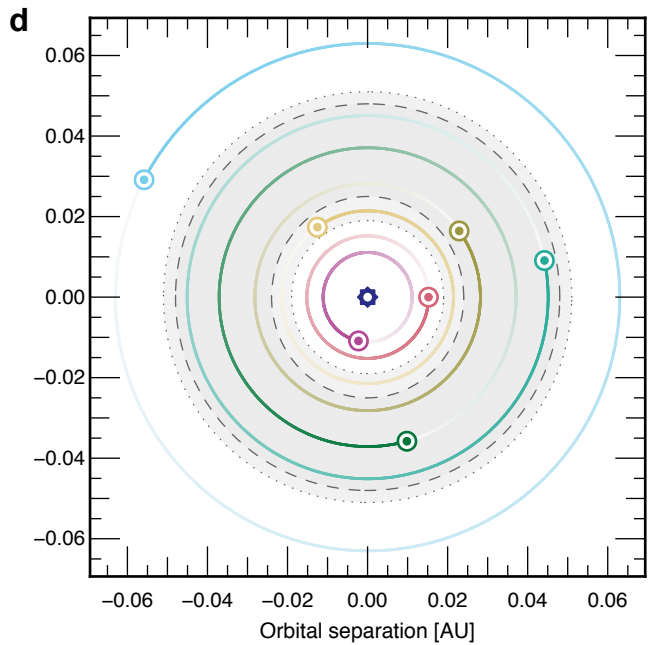

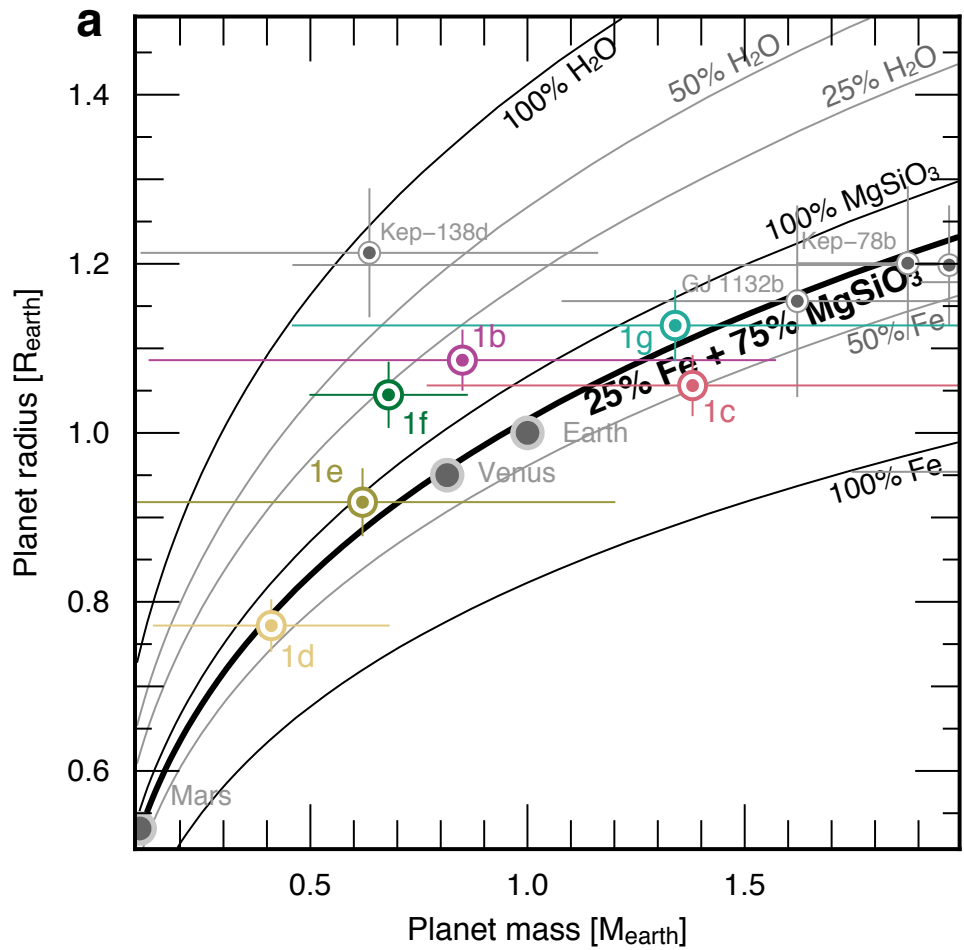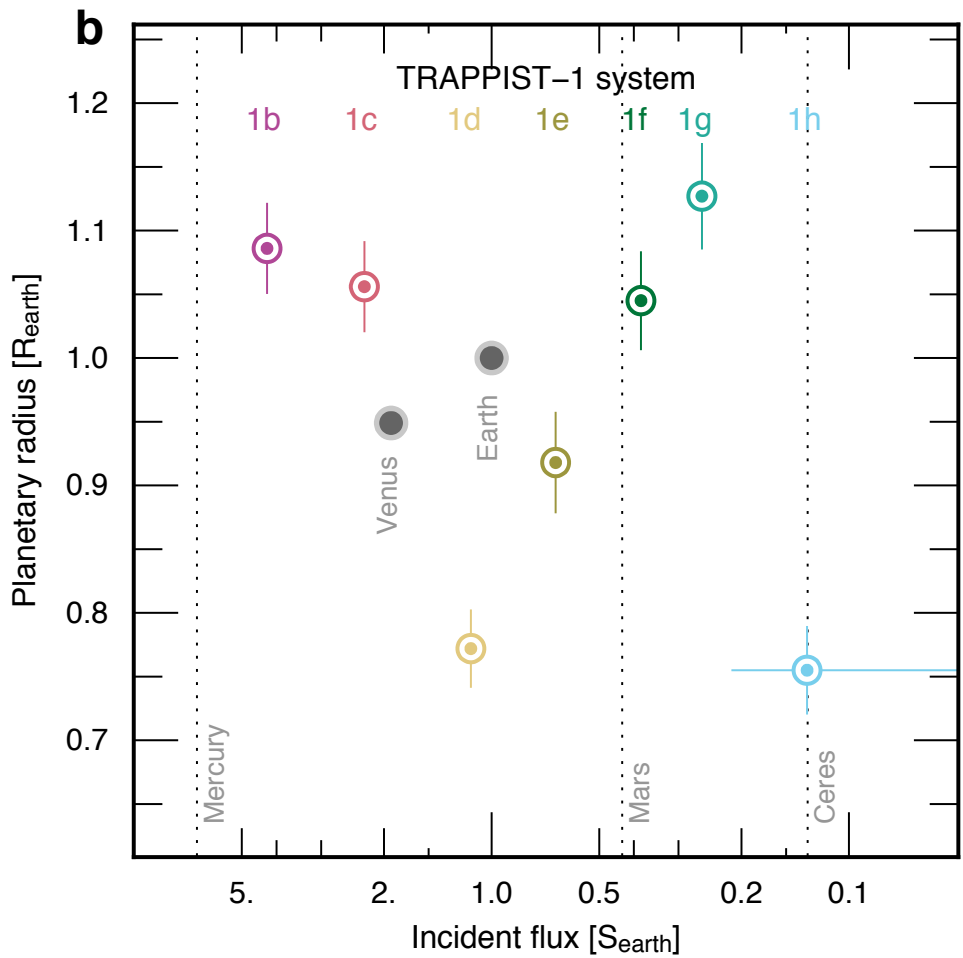

| Facility/instrument | Number of hrs | Year(s) | Number of light curves | Filter/grism | Number of transits |
|---|---|---|---|---|---|
| TRAPPIST-South | 677.9 | 2013 2015 2016 | 214 | I+z | b: 13, c: 1, d: 3, e: 5, f: 3, g: 4 |
| Spitzer/IRAC | 476.8 | 2016 | 30 | 4.5 μm | b: 16, c: 11, d: 5, e: 2, f: 3, g: 2, h: 1 |
| TRAPPIST-North | 206.7 | 2016 | 75 | I+z | b: 4, c: 3, e: 1 |
| LT/IO:O | 50.3 | 2016 | 10 | z' | b: 1, c: 1, e: 1, f: 1 |
| UKIRT/WFCAM | 34.5 | 2015 2016 | 9 | J | b: 4, c: 3 |
| WHT/ACAM | 25.8 | 2016 | 4 | I | b: 1, c: 1, d: 1 |
| SAAO-1m/SHOC | 10.7 | 2016 | 5 | z' | None |
| VLT/HAWK-I | 6.5 | 2015 | 2 | NB2090 | b: 1, c: 1, e: 1, f: 1 |
| HCT/HFOSC | 4.8 | 2016 | 1 | I | b: 1 |
| HST/WFC3 | 3.9 | 2016 | 1 | G141 (1.1-1.7 μm) | b: 1, c: 1 |

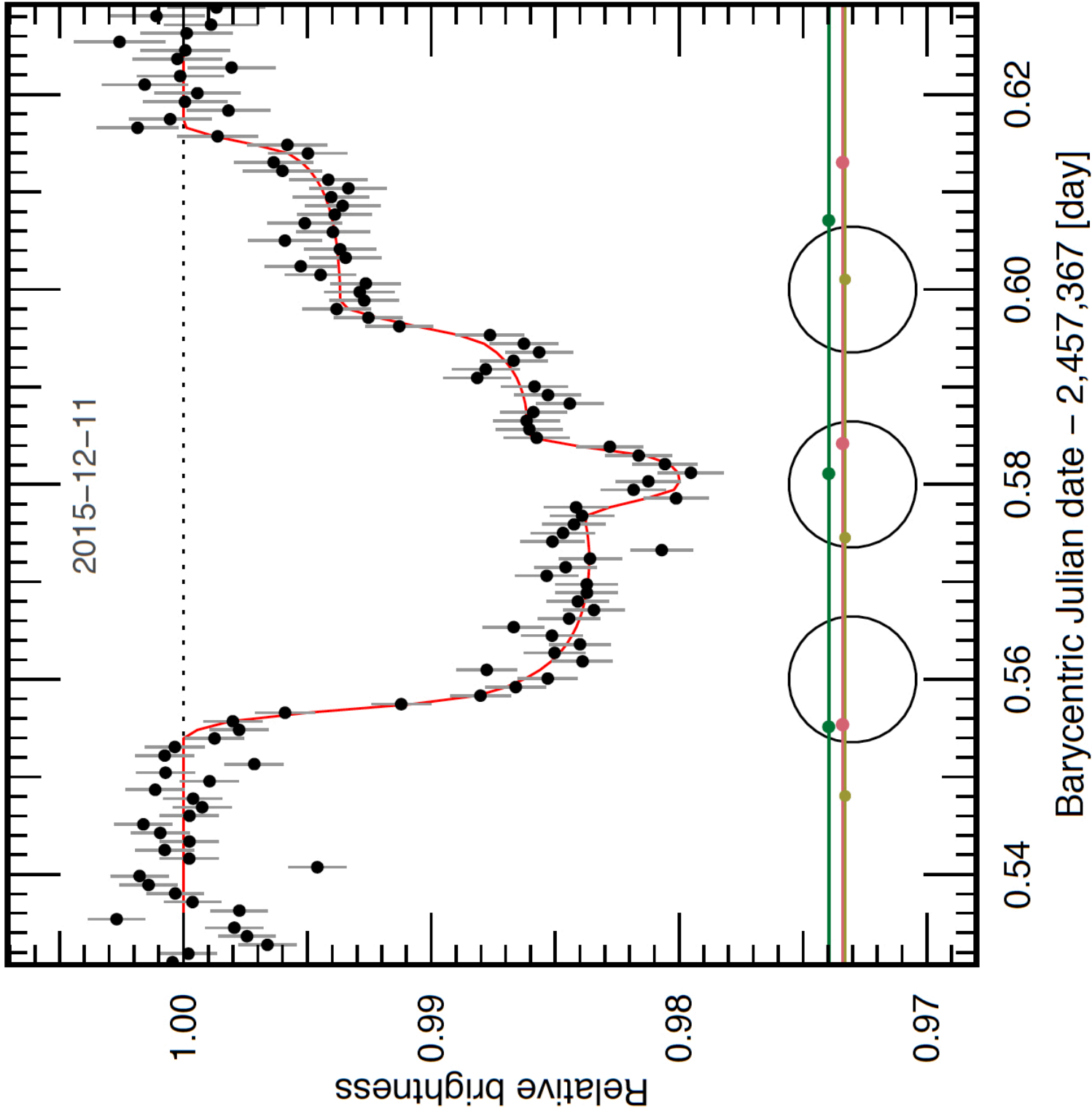

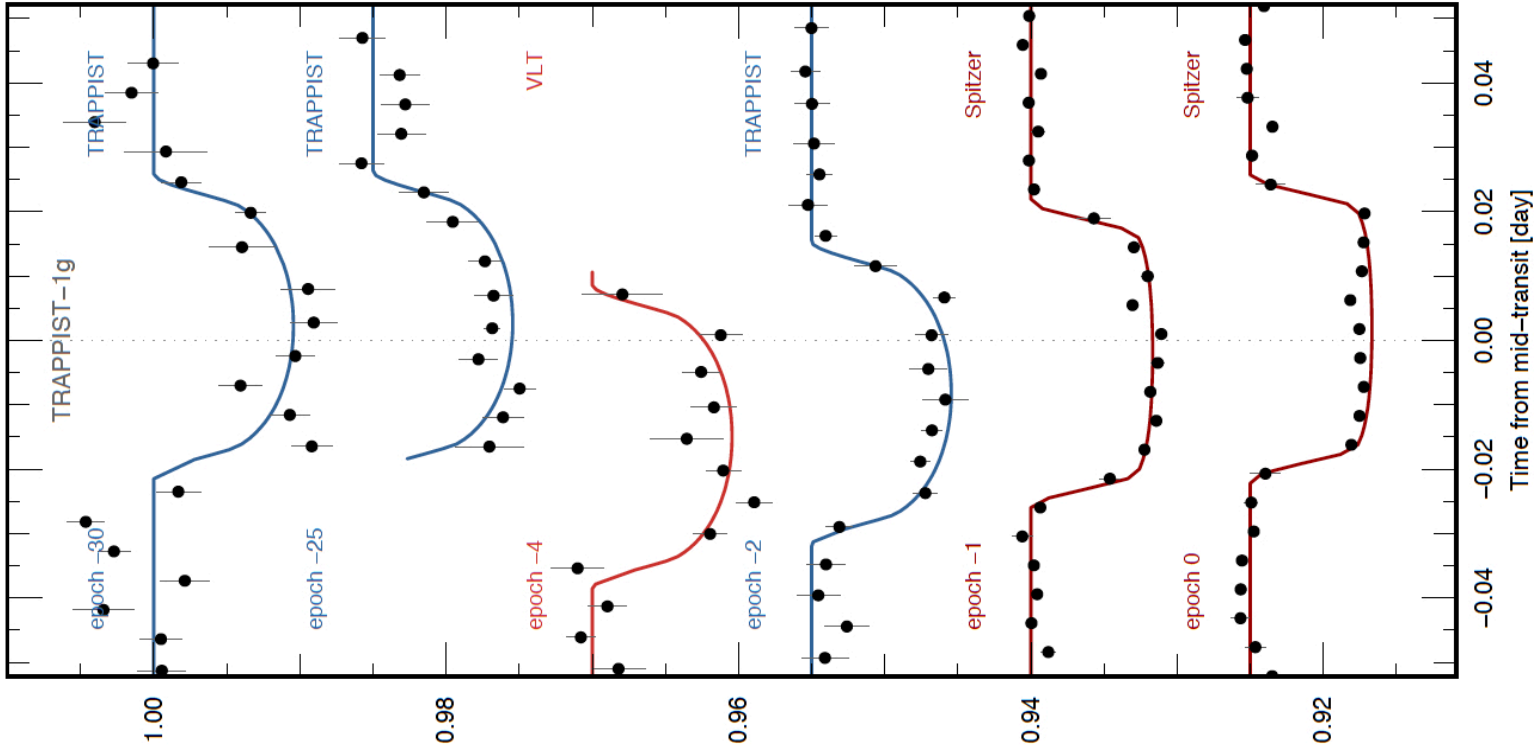
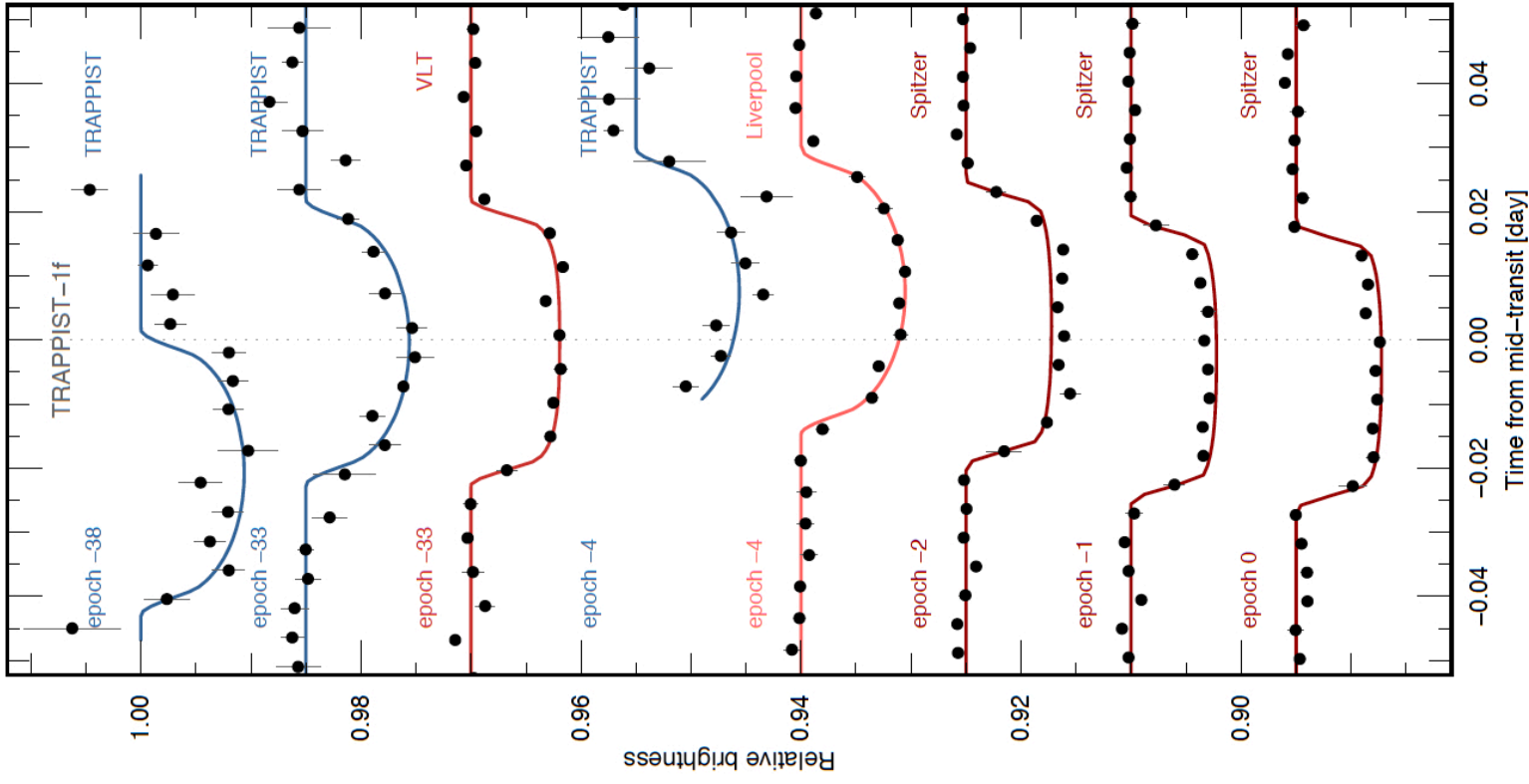

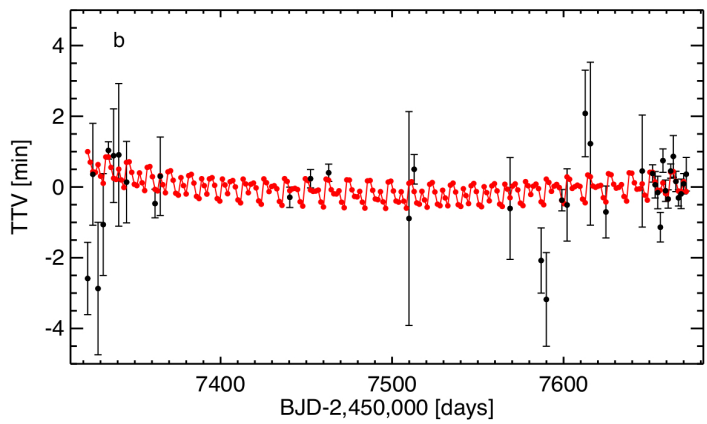
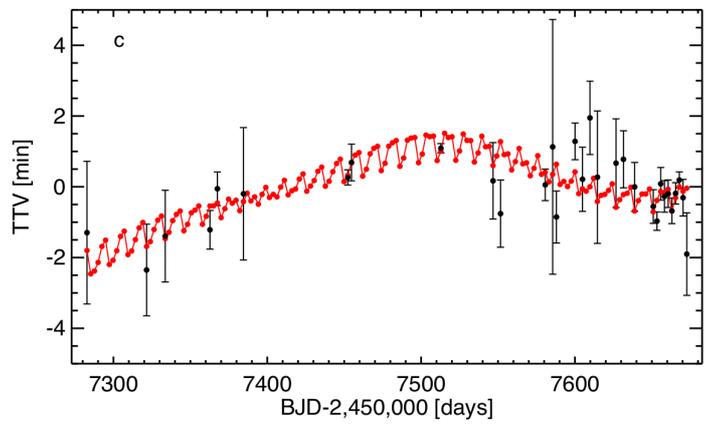
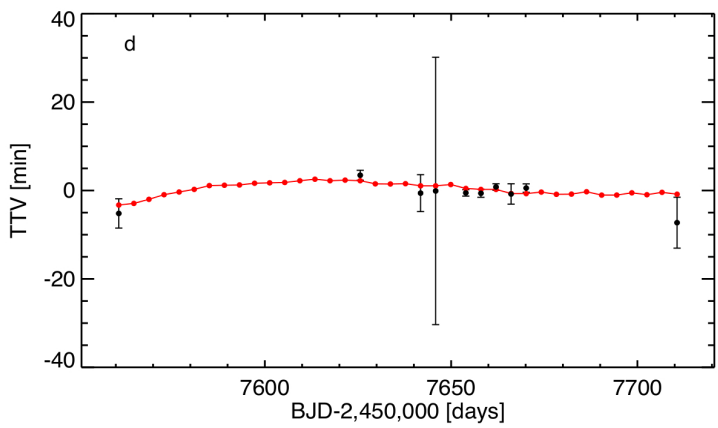
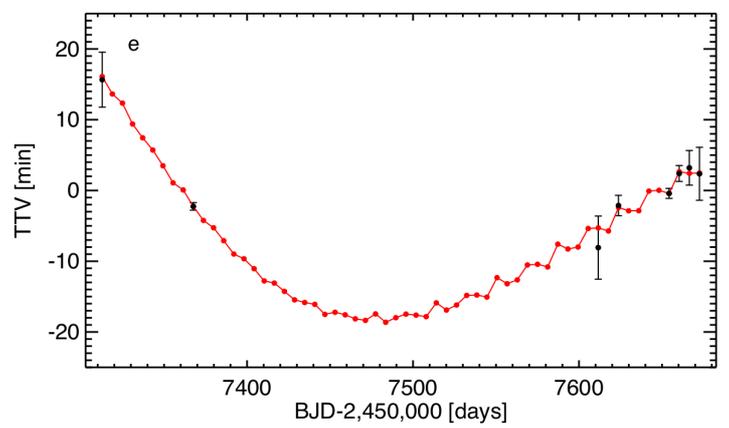
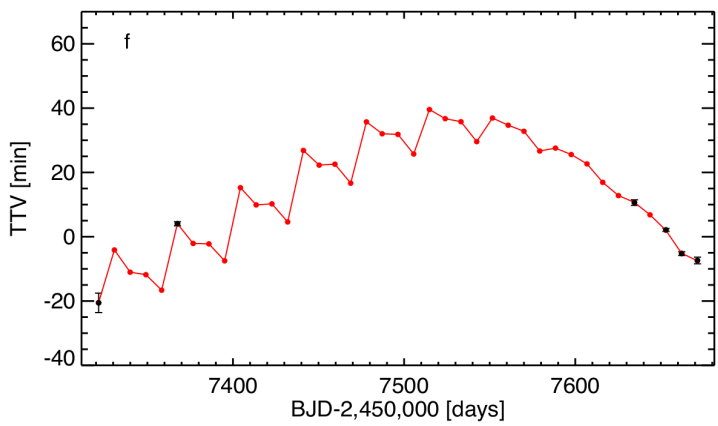
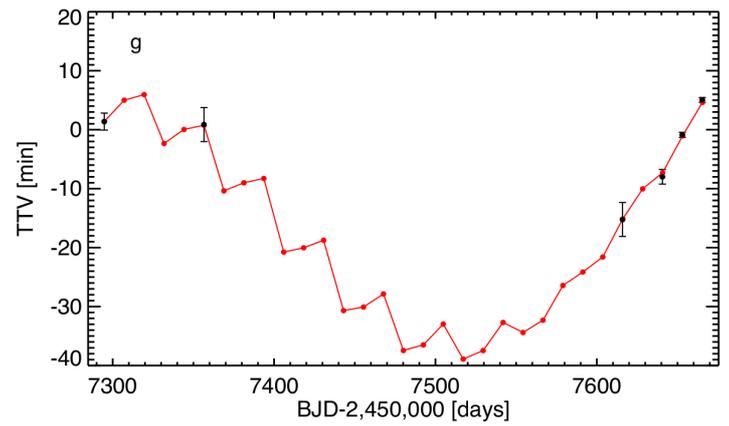

**Acknowledgments.** This work is based in part on observations made with the Spitzer Space Telescope, which is operated by the Jet Propulsion Laboratory, California Institute of Technology under a contract with NASA. The material presented here is based on work supported in part by NASA under Contract No. NNX15AI75G. TRAPPIST-South is a project funded by the Belgian F.R.S.-FNRS under grant FRFC 2.5.594.09.F, with the participation of the Swiss FNS. TRAPPIST-North is a project funded by the University of Liège, and performed in collaboration with Cadi Ayyad University of Marrakesh. The research leading to these results has received funding from the European Research Council under the FP/2007-2013 ERC Grant Agreement n° 336480 and under the H2020 ERC Grant Agreement n° 679030, and from the ARC grant for Concerted Research Actions, financed by the Wallonia-Brussels Federation. UKIRT is supported by NASA and operated under an agreement among the University of Hawaii, the University of Arizona, and Lockheed Martin Advanced Technology Center; operations are enabled through the cooperation of the East Asian Observatory. The Liverpool Telescope is operated on the island of La Palma by Liverpool John Moores University (JMU) in the Spanish Observatorio del Roque de los Muchachos of the Instituto de Astrofisica de Canarias with financial support from the UK Science and Technology Facilities Council. This paper uses observations made at the South African Astronomical Observatory (SAAO). MG, EJ, and VVG are F.R.S.-FNRS Research Associates. BOD acknowledges support from the Swiss National Science Foundation in the form of a SNSF Professorship (PP00P2_163967). EA acknowledges support from National Science Foundation (NSF) grant AST-1615315, and NASA grants NNX13AF62G and NNH05ZDA001. EB acknowledges that this work is part of the F.R.S.-FNRS ExtraOrDynHa research project and acknowledges funding by the European Research Council through ERC grant SPIRE 647383. SNR thanks the Agence Nationale pour la Recherche for support via grant ANR-13-BS05-0003-002 (project MOJO). DHL acknowledges financial support form the STFC. The authors thank C. Owen, C. Wolf, and the rest of the SkyMapper team for their attempts to monitor the star from Australia; for UKIRT the director R. Green and the staff scientists W. Varricatt and T. Kerr; the ESO staff at Paranal for their support on the HAWK-I observations; JMU and their flexibility for the LT schedule which allowed us to search actively for the planets, and to extend our time allocation in the face of amazing results; for the WHT, C. Fariña, F. Riddick, F. Jímenez and O. Vaduvescu for their help and kindness during observations; and for SAAO the telescopes operations manager R. Sefako for his support.